\documentclass[conference]{IEEEtran}
\usepackage{epsfig}
\usepackage{times}
\usepackage{float}
\usepackage{afterpage}
\usepackage{amsmath}
\usepackage{amstext}
\usepackage{amssymb,bm}
\usepackage{latexsym}
\usepackage{color}
\usepackage{graphicx}
\usepackage{amsmath}
\usepackage{amsthm}
\usepackage{graphicx}
\usepackage[center]{caption}
\usepackage{pstricks}
\usepackage{caption}
\usepackage{subcaption}
\usepackage{booktabs}
\usepackage{multicol}
\usepackage{lipsum}
 \usepackage[T1]{fontenc}
\usepackage{epstopdf}

\allowdisplaybreaks

\newtheorem{thm}{Theorem}

\newtheorem{rem}{Remark}
\newtheorem{example}{Example}
\theoremstyle{definition}

\providecommand{\definitionname}{Definition}

\newfloat{algorithm}{tbp}{loa}
\providecommand{\algorithmname}{Algorithm}
\floatname{algorithm}{\protect\algorithmname}

\usepackage{changes}
\definechangesauthor[color=blue,name={Mingyue Ji}]{MJ}

\long\def\comment#1{}

\newcommand{\dv}{{\mathbf d}}

\newcommand{\Zm}{{\mathbf Z}}

\newcommand{\Gc}{{\mathcal G}}
\newcommand{\Hc}{{\mathcal H}}

\newcommand{\Jc}{{\mathcal J}}

\newcommand{\Rc}{{\mathcal R}}

\newcommand{\Uc}{{\mathcal U}}
\newcommand{\Wc}{{\mathcal W}}

\newcommand{\Xc}{{\mathcal X}}

\newcommand{\Zc}{{\mathcal Z}}

\newcommand{\rsf}{{\mathsf r}}
\newcommand{\ssf}{{\mathsf s}}

\newcommand{\usf}{{\mathsf u}}

\newcommand{\Bsf}{{\mathsf B}}

\newcommand{\Hsf}{{\mathsf H}}

\newcommand{\Ksf}{{\mathsf K}}

\newcommand{\Msf}{{\mathsf M}}
\newcommand{\Nsf}{{\mathsf N}}

\newcommand{\Rsf}{{\mathsf R}}
\newcommand{\Ssf}{{\mathsf S}}

\renewcommand{\arg}{{\hbox{arg}}}

\newcommand{\be}{\begin{equation}}
\newcommand{\ee}{\end{equation}}
\newcommand{\bea}{\begin{eqnarray}}
\newcommand{\eea}{\end{eqnarray}}

\begin{document}

\title{On Combination Networks with Cache-aided Relays and Users} 


\author{
\IEEEauthorblockN{%
Kai~Wan\IEEEauthorrefmark{1},
Daniela~Tuninetti\IEEEauthorrefmark{3}
Pablo~Piantanida\IEEEauthorrefmark{1},
Mingyue~Ji\IEEEauthorrefmark{2},
}
\IEEEauthorblockA{\IEEEauthorrefmark{1}L2S CentraleSup\'elec-CNRS-Universit\'e Paris-Sud, Gif-sur-Yvette  91190, France, \{kai.wan, pablo.piantanida\}@l2s.centralesupelec.fr}%
\IEEEauthorblockA{\IEEEauthorrefmark{2}University of Utah, Salt Lake City, UT 84112, USA, mingyue.ji@utah.edu}%
\IEEEauthorblockA{\IEEEauthorrefmark{3}University of Illinois at Chicago, Chicago, IL 60607, USA, danielat@uic.edu}%
}

\maketitle
\begin{abstract}
Caching is an efficient way to reduce peak hour network traffic  congestion by storing some contents at the user's cache without knowledge of later demands. Coded caching strategy was originally proposed by Maddah-Ali and Niesen to give an additional coded caching gain compared the conventional uncoded scheme. Under  practical consideration, the caching model was recently considered in relay network, in particular the {\it combination network}, where  the  central server communicates with $\Ksf=\binom{\Hsf}{\rsf}$ users (each is with a cache of $\Msf$ files) through $\Hsf$ immediate relays, and each user is connected to a different $\rsf-$subsets of relays. Several inner bounds and outer bounds were proposed for combination networks with end-user-caches.  
This  paper extends the recent work by the authors on centralized combination networks with end-user caches to a more general setting, where both relays and users have caches. In contrast to the existing schemes in which the packets transmitted from the server are independent of the cached contents of relays, we propose a novel caching scheme by creating an additional coded caching gain to the transmitted load from the server  with the help of the cached contents  in relays. We also show that the proposed scheme outperforms the state-of-the-art approaches.
\end{abstract}

\section{Introduction}

\subsection{Shared Link Networks}
Caching content at the end-user's memories mitigates peak-hour network traffic congestion.
The seminal paper~\cite{dvbt2fundamental} by Maddah-Ali and Niesen (MAN) proposed an information theoretic model for cache-aided {\it shared link} networks. 
Such a model comprises a server with $\Nsf$ files of $\Bsf$ bits each, $\Ksf$ users with local memory of size $\Msf$ files, and a single error-free broadcast ``bottleneck'' link. A caching scheme comprises two phases. 
(i) {\it Placement phase}: during off-peak hours, the server places parts of its library into the users' caches without knowledge of what the users will later demand. 
{\it Centralized} caching systems allow for coordination among users during the placement phase, while {\it decentralized} ones do not. When pieces of files are simply copied into the cache, the cache placement phase is said to be {\it uncoded}; otherwise it is {\it coded}.
(ii) {\it Delivery phase}: each user requests one file during peak-hour time. 
According to the user demands and cache contents, the server transmits $\Rsf\Bsf$ bits  in order to satisfy all user demands. The goal is to determine $\Rsf^*$, the minimum load that satisfies {\it arbitrary}/worst-case user demands.

The {\it coded caching} strategy (with coded delivery) originally proposed in~\cite{dvbt2fundamental} gives an additional multiplicative {\it global caching gain} compared to uncoded caching schemes. For centralized systems, each file is split into a number of non-overlapping equal size and uncoded pieces that are  strategically placed  into the user caches. During the deliver phase, {\it coded multicast messages} are sent through the shared link so that a single transmission simultaneously serves several users. 
In~\cite{ontheoptimality}, we showed that the MAN scheme is optimal under the constraint of uncoded cache placement when $\Ksf \leq \Nsf$. In~\cite{exactrateuncoded}, the MAN scheme was shown  to have redundant multicast  transmissions when $\Ksf > \Nsf$. The achieved load in~\cite{exactrateuncoded} was proved to be optimal  under the constraint of uncoded placement, and that it is optimal to within a factor of 2~\cite{yas2}.

\begin{figure}
\centerline{\includegraphics[scale=0.16]{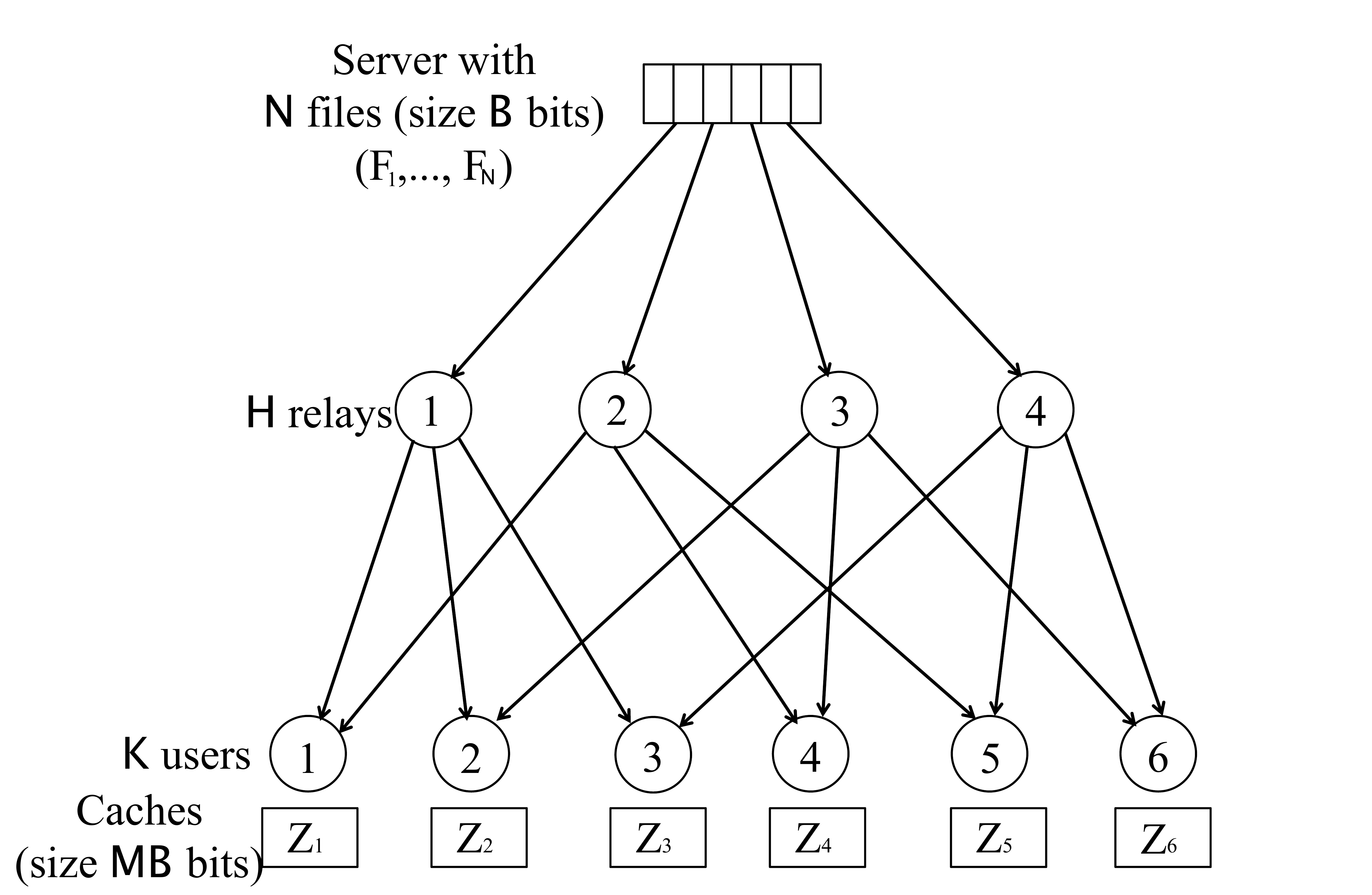}}
\caption{\small A combination network with end-user caches, with $\Hsf=4$ relays and $\Ksf=6$ users, i.e., $\rsf=2$.}
\label{fig: Combination_Networks}
\vspace{-5mm}
\end{figure}

\subsection{Combination Networks}
Since users may communicate with the central server through intermediate relays, recently caching  was considered in relay networks.  The caching problem with general relay networks was originally considered in~\cite{multiserver}, where a caching scheme including uncoded cache placement and linear network coding was proposed.
In~\cite{Naderializadeh2017onthoptimality}, it was proved that under the constraint of uncoded cache placement and of the separation between caching and multicast message generation on one hand, and message delivery on the other hand (i.e., the generation of the multicast messages is independent of the communication network topology), the proposed  scheme is order optimal within a factor of  $24$ among the separation schemes.

Since it is difficult to analyze  general relay networks, a symmetric network, known as {\it combination network}, has received a significant attention~\cite{cachingJi2015}. In this network as illustrated in Fig.~\ref{fig: Combination_Networks}, a server equipped with $\Nsf$ files is connected to $\Hsf$ relays. Each of the $\binom{\Hsf}{\rsf}$ users, with a memory of $\Msf$ files each, is connected to a unique $\rsf$-subset of  $\Hsf$ relays.  The goal is to minimize the maximum load among all links, which are assumed to be error-free and orthogonal.

The existing achievable schemes for centralized combination networks could be divided into two classes, based on uncoded cache placement~\cite{cachingincom,novelwan2017,wan2017novelmulticase,PDA2017yan} and cache placement~\cite{Zewail2017codedcaching,asymmetric2018wan,Wan2018ITA}, respectively. 
The authors in~\cite{cachingincom,novelwan2017} proposed delivery schemes to deliver MAN multicast messages. With MAN placement, we proposed a delivery by generating multicast messages based on the network topology in~\cite{wan2017novelmulticase}. The caching scheme proposed in~\cite{PDA2017yan} used the Placement Delivery Array (PDA) to reduce the sub-packetization of the schemes in~\cite{cachingincom,Zewail2017codedcaching} for the case $\rsf$ divides $\Hsf$.
The combination network was treated as $\Hsf$ uncoordinated shared-link models in~\cite{Zewail2017codedcaching} by using an $(\Hsf,\rsf)$ MDS precoding. By leveraging the connectivity of $\Ksf$ users and $\Hsf$ relays respectively, two asymmetric coded placements were proposed in~\cite{asymmetric2018wan,Wan2018ITA} which can lead a symmetric delivery phase. Outer bounds (based on cut-set or acyclic directed graph for the corresponding index coding problem) were proposed in~\cite{cachingJi2015,novelwan2017}.  
Some existing schemes are known to be optimal under the constraint of uncoded placement for some system parameters~\cite{novelwan2017,wan2017novelmulticase,asymmetric2018wan,PDA2017yan}.

\subsection{Beyond Combination Networks}
The existing inner and outer bounds for combination networks with end-user caches to more general settings:
\begin{enumerate}
\item Combination networks with cache-aided relays and users was considered in~\cite{Zewail2017codedcaching,wan2017novelmulticase}, where the main idea of~\cite{Zewail2017codedcaching,wan2017novelmulticase} is to divide each file into two parts, the part only cached in relays and the remaining part. The first parts of the demanded files are directly transmitted from relay to user and the delivery of the second parts is equivalent to the combination network with end-user-caches. 
\item 
The proposed scheme for centralized systems in~\cite{wan2017novelmulticase} was extended to decentralized systems with {dMAN} cache placement.
\item
In~\cite{wan2017novelmulticase}, we extended the proposed inner bound to more general  relay networks than combination networks, where each user is connected to an arbitrary subset of relays.
\end{enumerate}
\subsection{Contributions}
In this paper, we consider  combination networks with cache-aided relays and users, based on the asymmetric coded placement in~\cite{Wan2018ITA}, 
we propose a caching placement strategy where the cached contents  in relays are treated as the additional side informations of the connected users which can also help users to decode the coded messages transmitted from the server and thus
can further reduce  the transmitted load from the server to relays.  We also show that the proposed scheme outperforms the state-of-the-art schemes.

\section{System Model and Related Results}
\label{sec:system model}
In Section~\ref{sub:notation}, we introduce the notation convention used in this paper. In Section~\ref{sub:system model},  we introduce the system model for combination network with cache-aided relays and users. Finally, in Section~\ref{sub:our previous scheme}, we revise the asymmetric coded placement proposed by us in~\cite{Wan2018ITA}, which will be used in our proposed scheme for combination networks with cache-aided relays and users.
\subsection{Notation Convention}
\label{sub:notation}
A collection is a set of sets, e.g., $\big\{\{1,2\},\{1,3\} \big\}$.
Calligraphic symbols denote sets or collections, 
bold symbols denote vectors, 
and sans-serif symbols denote system parameters.
We use $|\cdot|$ to represent the cardinality of a set or the absolute value of a real number;
$[a:b]:=\left\{ a,a+1,\ldots,b\right\}$ and $[n] := [1:n]$; 
$\oplus$ represents bit-wise XOR. 
We define the set
\begin{align}
\arg \max_{x\in \Xc}f(x) := \big\{x\in \Xc:f(x)=\max_{y\in\Xc}f(y)\big\}.
\label{eq:def of argmaxfunct}
\end{align}
We define that 
\begin{align}
\Ksf_{i}:=\binom{\Hsf-i}{\rsf-i}, \ i\in [0:\rsf],
\label{eq:def of Ki}
\end{align}
where 
$\Ksf_0=\Ksf$ is the number of users in the system,
$\Ksf_1$ is the number of users connected to each relay,
and $\Ksf_i$ represents the number of users that are simultaneously connected to $i$ relays.
Our convention is that $\binom{x}{y}=0$ if $x<0$ or $y<0$ or $x<y$. 

\subsection{System Model for Combination Networks with Cache-aided Relays and Users}
\label{sub:system model}
In a $(\Hsf,\rsf,\Msf^{\textrm{relay}},\Msf^{\textrm{user}},\Nsf)$ combination network, a server has $\Nsf$ files, denoted by $F_1, \cdots, F_\Nsf$, each composed of $\Bsf$ i.i.d uniformly distributed bits.
The server is connected to $\Hsf$ relays through $\Hsf$ error-free orthogonal links. 
The relays are connected to $\Ksf := \Ksf_0$ users 
through $\rsf \, \Ksf$ error-free orthogonal links.
Each user  is connected to a distinct subset of $\rsf$ relays.  
Each relay can store $\Msf^{\textrm{relay}}\Bsf$ bits and each user can store $\Msf^{\textrm{user}}\Bsf$ bits, for $(\Msf^{\textrm{relay}},\Msf^{\textrm{user}})\in[0,\Nsf]^2$.
The subset of users connected to  relay $h$ is denoted by $\Uc_{h}, \ h\in[\Hsf]$. 
The subset of relays connected to user $k$ is denoted by $\Hc_{k}, k\in[\Ksf]$.
For each subset of users $\Wc\subseteq[\Ksf]$, the set of relays each of which is connected to all the users in $\Wc$ is denoted by 
\begin{align}
\Rc_{\Wc}:=\{h\in[\Hsf]: \Wc \subseteq \Uc_h \}.
\label{eq:RW def}
\end{align}
For the network in~Fig.~\ref{fig: Combination_Networks}, for example, 
$\Uc_{1}=\{1,2,3\}$,
$\Hc_{1}=\{1,2\}$, and 
$\Rc_{\{1,2\}}=\{1\}$.

In the placement phase, relay $h\in[\Hsf]$ and user $k\in[\Ksf]$ store information about the $\Nsf$ files in its cache of size $\mathsf{\Msf^{\textrm{relay}}\Bsf}$ and  $\mathsf{\Msf^{\textrm{user}}\Bsf}$ bits, respectively.
The cache content of relay $h\in[\Hsf]$ is denoted by $Z^{\textrm{relay}}_{h}$ and the one of
user $k\in[\Ksf]$ is denoted by $Z^{\textrm{user}}_{k}$; let $\Zm:=(Z^{\textrm{relay}}_{1},\ldots,Z^{\textrm{relay}}_{\Hsf},Z^{\textrm{user}}_{1},\ldots,Z^{\textrm{user}}_{\Ksf})$.
During the delivery phase, user $k\in[\Ksf]$ requests file $d_{k}\in[\Nsf]$;
the demand vector $\dv:=(d_{1},\ldots,d_{\Ksf})$ is revealed to all nodes. 
Given $(\dv,\Zm)$, the server sends a message $X_{h}$ 
of $\Bsf \, \Rsf_{h}(\dv,\Zm)$ bits to relay $h\in [\Hsf]$. 
Then, relay $h\in [\Hsf]$ transmits a message $X_{h\to k}$ 
of $\Bsf \, \Rsf_{h\to k}(\dv,\Zm)$ bits to user $k \in \Uc_h$. 
User $k\in[\Ksf]$ must recover its desired file $F_{d_{k}}$ from $Z_{k}$ and $(X_{h\to k} : h\in \Hc_k)$ with high probability when $\Bsf\to \infty$. 
The objective is to determine the load (number of transmitted bits in the delivery phase) pairs 
\[
\left(\Rsf^{\ssf\to \rsf},\Rsf^{\rsf\to \usf}\right) = 
\big(
\max_{h\in[\Hsf]}\Rsf_h(\dv,\Zm), 
\max_{h\in[\Hsf],k\in \Uc_h}\Rsf_{h\to k}(\dv,\Zm)
\big)
\]
for the worst case demands $\dv$ for a given placement $\Zm$.

In practice,  the throughput of transmission from the server to relays may be much lower than the throughput from the relays to their local connected users. For example, in wireless networks where the throughput from small cell base stations to users are much higher than that from the macro base stations to small base stations if all use sub-6GHz wireless communications. 
In this paper, for combination networks with cache-aided relays and users, we mainly want to minimize the max-link load from the server to relays, i.e., $\Rsf^{\ssf\to \rsf}=\max_{h\in[\Hsf]}\Rsf_h(\dv,\Zm)$.

For a  caching scheme with max-link load among all the links from the server to relays $\Rsf^{\ssf\to \rsf}$, we say it attains a  {\it coded caching gain} of $g$ if
\begin{align}
&\Rsf^{\ssf\to \rsf}
 = \frac{\Rsf^{\ssf\to \rsf}_{\text{routing}}}{g}, \ \text{for} \\
&\Rsf^{\ssf\to \rsf}_{\text{routing}} := \frac{\Ksf\max \big\{1-(\rsf\Msf^{\textrm{relay}}+\Msf^{\textrm{user}})/\Nsf,0\big\}}{\Hsf} \nonumber\\
 & = \frac{\Ksf_1\max\big\{1-(\rsf\Msf^{\textrm{relay}}+\Msf^{\textrm{user}})/\Nsf,0\big\}}{\rsf},
\end{align}
where $\Rsf^{\ssf\to \rsf}_{\text{routing}}$ is achieved by routing.
By the cut-set bound~\cite{cachingincom} we have $g\leq \Ksf_1$ (recall that $\Ksf_1$ is the number of users connected to each relay).

\subsection{Asymmetric Coded Placement in~\cite{Wan2018ITA}}
\label{sub:our previous scheme}
In this part, we introduce the caching scheme based on an asymmetric coded placement for the case $\Msf^{\textrm{relay}}=0$ proposed in~\cite{Wan2018ITA}, which treats the combination network as $\Hsf$ coordinated shared-link models and leverages the connectivity among the divided models.

 We aim to achieve coded caching gain $g\in[2:\Ksf_1]$, that is, every coded multicast message is simultaneously useful for $g$ users. So each subfile should be cached by at least $g-1$  users.
We consider each subset of users $\Wc$ with cardinality $g-1$ for which there exists at least one relay connected to all the users in $\Wc$,
that is, we define the collection 
\begin{align}
\Zc_g:=\big\{\Wc\subseteq [\Ksf]:|\Wc|=g-1,\Rc_{\Wc}\neq\emptyset\big\},
\label{eq:def of Z}
\end{align}
where $\Rc_{\Wc}$ defined in~\eqref{eq:RW def}.  For example, consider the combination network in Fig.~\ref{fig: Combination_Networks}, we have 
\begin{align*}
&\Zc_3=\big\{\{1,2\},\{1,3\},\{1,4\},\{1,5\},\{2,3\},\{2,4\},\{2,6\},\\
&\{3,5\},\{3,6\},\{4,5\},\{4,6\},\{5,6\} \big\}.
\end{align*}
Each subfile corresponds to one set in $\Zc_g$.

\paragraph*{Placement phase}
 We define 
\begin{align}
&\Ssf_1(g):=\sum_{a=1}^{\rsf}\binom{\rsf}{a}\binom{\Ksf_a}{g-1}(-1)^{a-1}.\label{eq:def of S1}\\
&\Ssf_2(g):=\sum_{a=1}^{\rsf}\binom{\rsf}{a}\binom{\Ksf_a-1}{g-2}(-1)^{a-1} \label{eq:def of S2}.
\end{align}
We divide each file $i\in[\Nsf]$ into 
$\Ssf_1(g)$ non-overlapping and equal-length pieces, which are then encoded by a  $(|\Zc_g|,\Ssf_1(g))$ MDS code;
denote the MDS coded symbols/subfiles as $(f_{i,\Wc} : \Wc\in \Zc_g)$.
For each $\Wc\in \Zc_g$ 
$f_{i,\Wc}$ is cached by the users in $\Wc$. 
Each MDS coded symbol includes $\frac{\Bsf}{\Ssf_1(g)}$ bits,  and  thus  by the inclusion-exclusion principle~\cite[Theorem~10.1]{combinatorics}, we can compute that  the needed memory size is 
\begin{align}
\Msf^{\textrm{user}}_\text{\rm b}
=\frac
{\Nsf\Ssf_2(g)}
{\Ssf_1(g)}.
\label{eq:Mb def}
\end{align}
\paragraph*{Delivery phase}
Each user $k\in[\Ksf]$ needs to recover all the MDS coded symbols $f_{d_k,\Wc}$ where $\Wc\in\Zc_g$, $k\notin \Wc$ and $\Rc_{\{k\}\cup\Wc}\neq \emptyset$ (but not those for which $\Rc_{\{k\}\cup\Wc}= \emptyset$).
For those MDS coded symbols needed by user $k$, we divide $f_{d_k,\Wc}$ into $|\Rc_{\{k\}\cup\Wc}|$ non-overlapping and equal-length pieces, $f_{d_k,\Wc}=\{f_{d_k,\Wc,h}:h\in \Rc_{\{k\}\cup\Wc}\}$.  After considering all the MDS coded symbols demanded by all the users, 
for each relay $h\in[\Hsf]$ and each set $\Jc\subseteq \Uc_h$ where $|\Jc|=g$, we create the multicast message 
\begin{align}
W_{\Jc}^{h}:=\underset{k\in\Jc}{\oplus}f_{d_k,\Jc\setminus \{k\},h}
\label{eq:SRDS multicast messages}
\end{align}
which will be sent to relay $h$ who will then forward it to the users in $\Jc$. 

Hence from the placement and delivery phase, each user $k\in[\Ksf]$ obtains the MDS coded symbol $f_{d_k,\Wc}$ where $\Wc\in\Zc_g$ and $\Rc_{\{k\}\cup\Wc}\neq \emptyset$. By  the inclusion-exclusion principle~\cite[Theorem~10.1]{combinatorics}, user $k$ totally obtains $\Ssf_1(g)$ MDS coded symbols of $F_{d_k}$ such that it can recover its desired file $F_{d_k}$.
\paragraph*{Max-link load} 
It can be proved that
each demanded MDS coded symbol is multicasted with other $g-1$ demanded MDS coded symbols with the same length and thus the coded caching gain is $g$ and thus the max link-load is $\Ksf(1-\Msf^{\textrm{user}}_\text{\rm b}/\Nsf)/(\Hsf g)$.

It was also shown in~\cite{Wan2018ITA} that when $g-1\leq \Ksf_2$, the achieved max link-load by the proposed approach is strictly lower than the one by~\cite{Zewail2017codedcaching}. However,
when $g-1>\Ksf_2$, we have $|\Rc_\Wc|=1$ for each set $\Wc\in \Zc_g$, and thus
 we do not leverage the coordination among relays. In this case it is equivalent to the scheme in~\cite{Zewail2017codedcaching}. 
\section{Combination Networks with Cache-aided Relays and Users}
\label{sec:cache-aided relays and users}

In Section~\ref{sub:Yeners scheme} we will revise the caching scheme in~\cite{Zewail2017codedcaching}, which divides each file into two parts and the packets transmitted from the server are independent of the cached contents of relays. In Section~\ref{sub:proposed scheme cache aided relays}, we propose a novel caching scheme, in which the users can leverage the cached contents of the connected relays to decode the coded messages transmitted from the server. 
\subsection{Caching in~\cite{Zewail2017codedcaching} for Combination Networks with Cache-aided Relays and Users}
\label{sub:Yeners scheme}
The memories-loads tradeoff of the scheme in~\cite{Zewail2017codedcaching} is the lower convex envelope of the two groups of points.
\begin{enumerate}
\item $(\Msf^{\textrm{relay}},\Msf^{\textrm{user}})=(0,\Nsf t_2/\Ksf_1)$ where $t_2\in [0:\Ksf_1]$. For each point in this group, we can see that relays do not have memory and the scheme is equivalent to the one for combination networks with end-user-caches. The combination network is treated as $\Hsf$ uncoordinated shared-link models. 
\paragraph*{Placement Phase}
Each file $F_i$, where $i\in[\Nsf],$ is divided into $\rsf$ non-overlapping and equal-length pieces, which are then encoded by using a $(\Hsf,\rsf)$ MDS code; the $h$-th MDS coded symbol is denoted by  $s^{h}_{i}$, of size $|s^{h}_{i}|=\Bsf/\rsf$ for $h\in[\Hsf]$.
For each $h\in [\Hsf]$,  
$s^{h}_{i}$ is divided into $\binom{\Ksf_1}{t_2}$ non-overlapping and equal-length pieces, i.e., $s^{h}_{i}=\{s^{h}_{i,\Wc}:\Wc\subseteq \Uc_h, |\Wc|=t_2\}$. 
Each user $k\in \Uc_{h}$ caches $s^{h}_{i,\Wc}$ if $k\in \Wc$. 
\paragraph*{Delivery Phase}
For each relay $h\in[\Hsf]$,
the MAN-like multicast messages
\begin{align}
w^{h}_{\Jc}=\underset{k\in\Jc}{\oplus}s^{h}_{ d_k,\Jc\setminus \{k\}} , 
\ \forall \Jc\subseteq \Uc_h : |\Jc|=t_2+1, \ h\in [\Hsf],
\end{align}
are delivered from the server to relay $h$, and then
relay $h$ then forwards $w^{h}_{\Jc}$ to the users in $\Jc$. 
It can be seen that user $k$ connected to relay $h$ can recover $s^{h}_{d_k}$, and eventually 
$F_{d_k}$.  
\paragraph*{Max-link load} 
Each demanded subfile is transmitted with other $t_2$ subfiles from the server. So $\Rsf^{\ssf\to \rsf}=\frac{\Ksf(1-\Msf^{\textrm{user}}/\Nsf}{\Hsf (t_2+1)})$. Each user receives the uncached part of its demanded file with totally $(1-\Msf^{\textrm{user}}/\Nsf)\Bsf$ bits from its  $\rsf$ connected relays. So 
$\Rsf^{\rsf\to \usf}=(1-\Msf^{\textrm{user}}/\Nsf)/\rsf$.
\item $(\Msf^{\textrm{relay}},\Msf^{\textrm{user}})=(\Nsf/\rsf,\Nsf t_1/\Ksf_1)$ where $t_1\in \{0,\Ksf_1\}$. In this case, each relay directly caches $s^h_{i}$ such that the server needs not to transmit any packets to relays. 

If $t_1=0$, each user does not cache any bits. In the delivery phase, each relay $h\in[\Hsf]$ transmits $s^h_{d_k}$ to each user $k\in\Uc_h$. So we have $(\Rsf^{\ssf\to \rsf},\Rsf^{\rsf\to \usf})=(0,1/\rsf)$.

If $t_1=\Ksf_1$, in the placement phase, each user $k$ caches $s^h_{i}$ for $h\in[\Hsf]$ and $i\in[\Nsf]$. So it caches all the $\Nsf$ files such that $(\Rsf^{\ssf\to \rsf},\Rsf^{\rsf\to \usf})=(0,0)$. 
\end{enumerate}
Notice that in each of the above points, $\Rsf^{\rsf\to \usf}$ is always equal to $(1-\Msf^{\textrm{user}}/\Nsf)/\rsf$.
\subsection{Proposed Scheme for Combination Networks with Cache-aided Relays and Users}
\label{sub:proposed scheme cache aided relays}
We start by an example of~\cite{Wan2018ITA} for combination networks with end-user-caches, which will be used later to derive our  proposed method for combination network with cache-aided relays and users.
\begin{example}[$\Hsf=5$, $\rsf=3$, $\Nsf=10$, $\Msf^{\textrm{relay}}=0$ and $g=3$]
\label{ex:end-cache-users}
In this example, we have
\begin{align*}
  &\Uc_{1}=[6], \ \Uc_{2}=\{1, 2, 3, 7, 8, 9\},\ \Uc_{3}=\{1, 4, 5, 7, 8, 10\},
\\&\Uc_{4}=\{2, 4, 6, 7, 9, 10\},\ \Uc_{5}=\{3, 5, 6, 8, 9, 10\}.
\end{align*}
We aim to achieve coded caching gain $g=3$, that is, every multicast message is simultaneously useful for $g=3$ users. So each subfile should be known by at least $g-1=2$ users.
\paragraph*{Placement phase}
We divide each $F_{i}$ into $\Ssf_1(g)=36$ non-overlapping and equal-length pieces, which are then encoded by  $(|\Zc_3|=45,36)$ MDS code. The length of each MDS symbol is $\Bsf/36$. For each $\Wc\in \Zc_3$, there is one MDS coded symbol/subfile  denoted by $f_{i,\Wc}$ (composed of bits) which  is cached by all the users in $\Wc$. Thus the memory size $\Msf^{\textrm{user}}=\Nsf\Ssf_2(g)/\Ssf_1(g)=2.5$.
\paragraph*{Delivery phase}
We let each user $k$ recover $f_{d_k, \Wc}$ where $k\notin \Wc$, $\Wc\in\Zc_g$ and $\Rc_{\{k\}\cup \Wc}\neq \emptyset$. For each such $f_{d_k,\Wc}$, we divide it into $|\Rc_{\{k\}\cup\Wc}|$ non-overlapping and equal-length pieces, $f_{d_k,\Wc}=\{f_{d_k,\Wc,h}:h\in \Rc_{\{k\}\cup\Wc}\}$.
After considering all the MDS coded symbol demanded by all the  users, 
for each relay $h\in[\Hsf]$ and each set $\Jc\subseteq \Uc_h$ where $|\Jc|=g$, we create the multicast message in~\eqref{eq:SRDS multicast messages} 
to be sent to relay $h$ and then forwarded to the users in $\Jc$. Hence each demanded MDS coded symbol  is transmit in one linear combination which also includes other $g-1$ demanded MDS coded symbols with identical length and thus the coded caching gain is $g=3$.
In conclusion, the minimum needed memory size to achieve $g=3$ of the proposed scheme is $\Msf^{\textrm{user}}=2.5$ while the ones of~\cite{Zewail2017codedcaching} is $10/3\approx 3.333$.
\end{example}
Our proposed scheme for combination networks with cache-aided relays and users illustrated in the next example is based on the caching scheme in Example~\ref{ex:end-cache-users}. 
\begin{example}[$\Hsf=5$, $\rsf=3$, $\Nsf=10$, $\Msf^{\textrm{relay}}=25/12$ and $\Msf^{\textrm{user}}=5/12$, $g=3$]
\label{ex:cache-aided relays and users}
The network topology is the same as Example~\ref{ex:end-cache-users}. In this example, we also impose that each demanded subfile of each user which is neither stored in its memory nor the memories of its connected relays, is transmitted from the server in one linear combination including other $g-1=2$ subfiles.
We aim to let each user benefit from the cached content in its connected relays as its own cache contents.
\paragraph*{Placement phase}
As in Example~\ref{ex:end-cache-users}, we also divide each $F_{i}$ into $36$ non-overlapping and equal-length pieces, which are then encoded by  $(45,36)$ MDS code. The length is each MDS symbol is $\Bsf/36$.
For each $\Wc\in \Zc_3$, there is one MDS symbol  denoted by $f_{i,\Wc}$. However, different to Example~\ref{ex:end-cache-users}, not the whole symbol $f_{i,\Wc}$ is stored in the cache of each user in $\Wc$. 
Instead, we divide $f_{i,\Wc}$ into $|\Rc_{\Wc}|+1$ non-overlapping parts   (but not necessary with identical length), $f_{i,\Wc}=\{f_{i,\Wc,h}:h\in \Rc_{\Wc}\}\cup\{f^{\prime}_{i,\Wc}\}$. For each $h\in \Rc_{\Wc}$, $f_{i,\Wc,h}$ is cached by relay $h$ where $|f_{i,\Wc,h}|=\frac{\Msf^{\textrm{relay}}\Bsf}{\Nsf\binom{\Ksf_1}{g-1}}$.  
In addition, $f^{\prime}_{i,\Wc}$ is cached by each user in $\Wc$ where
$|f^{\prime}_{i,\Wc}|=\Bsf/36-\frac{|\Rc_{\Wc}|\Msf^{\textrm{relay}}\Bsf}{\Nsf\binom{\Ksf_1}{g-1}}$.  
 Hence, each relay $h\in[\Hsf]$ caches $f_{i,\Wc,h}$ for each $\Wc\subseteq \Uc_h$ and each $i\in[\Nsf]$ where  $|\Wc|=g-1$. Thus the number of cached bits of relay $h$ is
$$
\Nsf\binom{\Ksf_1}{g-1}\frac{\Msf^{\textrm{relay}}\Bsf}{\Nsf\binom{\Ksf_1}{g-1}}=\Msf^{\textrm{relay}}\Bsf.
$$

For example, consider $f_{i,\{1,2\}}$ which is divided into $|\Rc_{\{1,2\}}|+1=3$ non-overlapping pieces. Each relay $h\in \Rc_{\{1,2\}}=\{1,2\}$ caches $f_{i,\{1,2\},h}$ with  $\frac{\Msf^{\textrm{relay}}\Bsf}{\Nsf\binom{\Ksf_1}{g-1}}=\Bsf/72$ bits. So for the last piece, we have $|f^{\prime}_{i,\{1,2\}}|=\Bsf/36-2\Bsf/72=0$ and thus no user caches any bits of $f_{i,\{1,2\}}$.

Consider now $f_{i,\{1,6\}}$ which is divided into $|\Rc_{\{1,6\}}|+1=2$ non-overlapping pieces. Each relay $h\in \Rc_{\{1,6\}}=\{1\}$ caches $f_{i,\{1,6\},h}$ with  $\frac{\Msf^{\textrm{relay}}\Bsf}{\Nsf\binom{\Ksf_1}{g-1}}=\Bsf/72$ bits. So $|f^{\prime}_{i,\{1,6\}}|=\Bsf/36-\Bsf/72=\Bsf/72$. Thus each user in $\{1,2\}$ caches $f^{\prime}_{i,\{1,6\}}$ with $\Bsf/72$ bits.


We then focus on user $1$. For each set $\Wc\in \big\{\{1,2\},\{1,3\},\{1,4\},\{1,5\}, \{1,7\},\{1,8\}\big\}$, we have $|\Rc_{\Wc}|=2$ and $|f^{\prime}_{i,\Wc}|=0$. For each set $\Wc \in \big\{\{1,6\},\{1,9\},\{1,10\} \big\}$, we have $|\Rc_{\Wc}|=1$ and $|f^{\prime}_{i,\Wc}|=\Bsf/72$. So user $1$  caches $3\Nsf\Bsf/72=\Msf^{\textrm{user}}\Bsf$ bits.
\paragraph*{Delivery phase}
We let each user $k$ recover $f_{d_k,\Wc}$ where $\Wc \in \Zc_3$ and $\Rc_{\{k\}\cup \Wc}\neq\emptyset$. There are three steps in delivery phase:
\begin{enumerate}
\item In the first step, for each relay $h\in [\Hsf]$ and each user $k\in \Uc_h$, relay $h$ delivers all the cached bits of $F_{d_k}$ to user $k$.  More precisely, for each set $\Wc\subseteq \Uc_h$ where $|\Wc|=g-1$, relay $h$ delivers $f_{d_k,\Wc,h}$ to user $k$.
So by this step and the placement phase, each user $k\in[\Ksf]$ can recover $f_{d_k,\Wc}$ where $\Wc \in \Zc_3$ and $k\in \Wc$. User $k$ can also recover $f_{d_k,\Wc,h}$ where  $\Wc \in \Zc_3$, $k\notin \Wc$, $\Rc_{\{k\}\cup\Wc}\neq \emptyset$ and $h\in (\Rc_{\Wc}\cap\Hc_k)$.

 \item In the second step, we also focus on each relay $h\in [\Hsf]$ and each user $k\in \Uc_h$.  For each set $\Wc^{\prime}\subseteq \Uc_h$ and each $k^{\prime}\subseteq [\Ksf]\setminus \Uc_h$ where $|\Wc^{\prime}|=g-1$, $k\in \Wc^{\prime}$ and $\Rc_{\{k^{\prime}\}\cup \Wc^{\prime}}\neq \emptyset$,
relay $h$ delivers $f_{d_{k^{\prime}},\Wc^{\prime},h}$ to user $k$. These additional side information of user $k$ will help him decode the multicast messages transmitted from the server in the second step.
\item In the last step, as Example~\ref{ex:end-cache-users}, we let each user $k$ recover $f_{d_k,\Wc}\setminus\{f_{d_k,\Wc,h}:h\in (\Rc_{\Wc}\cap\Hc_k)\}$ where  $\Wc \in \Zc_3$, $k\notin \Wc$ and $\Rc_{\{k\}\cup\Wc}\neq \emptyset$. 
More precisely, we let $\Gc_{k,\Wc}=f_{d_k,\Wc}\setminus\{f_{d_k,\Wc,h}:h\in (\Rc_{\Wc}\cap\Hc_k)\}$ representing the unknown bits in $f_{d_k,\Wc}$ of user $k$.  We divide $\Gc_{k,\Wc}$ into $|\Rc_{\{k\}\cup\Wc}|$ non-overlapping and equal-length pieces, $\Gc_{k,\Wc}=\{\Gc_{k,\Wc,h}:h\in \Rc_{\{k\}\cup\Wc}\}$.
After considering all the MDS coded symbols demanded by all the  users, 
for each relay $h\in[\Hsf]$ and each set $\Jc\subseteq \Uc_h$ where $|\Jc|=g=3$, we create the multicast message 
\begin{align}
V^{h}_{\Jc}=\underset{k\in\Jc}{\oplus}\Gc_{k,\Jc\setminus \{k\},h}\label{eq:proposed scheme multicast messages}
\end{align}
to be sent to relay $h$ and then forwarded to the users in $\Jc$, where we use the same convention as that in the literature  when it comes to `summing' sets. 

For example, consider relay $1$ and set $\Jc_1=\{1,2,3\}$. It can be seen that $\Gc_{1,\{2,3\}}=f_{d_1,\{2,3\}}\setminus(f_{d_1,\{2,3\},1}\cup f_{d_1,\{2,3\},2})=\emptyset$. Similarly $\Gc_{2,\{1,3\}}=\Gc_{3,\{1,2\}}=\emptyset$. So $V^{1}_{\{1,2,3\}}=\emptyset$.

Consider now relay $1$ and set $\Jc_2=\{1,2,4\}$. It can be seen that $\Gc_{1,\{2,4\}}=f_{d_1,\{2,4\}}\setminus f_{d_1,\{2,4\},1}=f_{d_1,\{2,4\},4}$. Since $\Rc_{\{1,2,4\}}=\{1\}$, we don't further partition  $\Gc_{1,\{2,4\}}$ which includes  $\Bsf/72$ bits. Similarly, each of $\Gc_{2,\{1,4\}}$ and $\Gc_{4,\{1,2\}}$ has $\Bsf/72$ bits. Hence, $V^{1}_{\{1,2,4\}}=\Gc_{1,\{2,4\}}\oplus\Gc_{2,\{1,4\}}\oplus \Gc_{4,\{1,2\}}$ 
including $\Bsf/72$ is transmitted from the server to relay $1$, which then forwards it to users in $\{1,2,4\}$.
\end{enumerate} 

Hence, we achieve $g=3$ and
 $(\Rsf^{\ssf\to \rsf},\Rsf^{\rsf\to \usf})=(\frac{2}{9},\frac{41}{72})\approx (0.22,0.57)$ while the scheme in~\cite{Zewail2017codedcaching} described in Section~\ref{sub:Yeners scheme} gives $(\Rsf^{\ssf\to \rsf},\Rsf^{\rsf\to \usf})=(\frac{11}{24},\frac{23}{72})\approx (0.46,0.32)$. It can be seen the max link-load from the server to relays achieved by the proposed method is less than the half of the one achieved by the scheme in~\cite{Zewail2017codedcaching}. 

Comparing the proposed scheme and the scheme in~\cite{Zewail2017codedcaching}, there are main two  advantages. On one hand, we can see that the cached contents of relays help users to decode the packets transmitted from the server which can lead an additional coded caching gain. For example, $f_{d_7,\{1,2\},1}$ is cached by relay $1$ and $f_{d_2,\{1,7\},3}$ is cached by relay $3$. In the first step of delivery, $f_{d_7,\{1,2\},1}$ is transmitted from relay $1$ to user $1$ and $f_{d_2,\{1,7\},3}$ is transmitted from relay $3$ to user $1$ such that user $1$ knows them. In the second step of delivery, the server transmit $f_{d_1,\{2,7\},4}\oplus f_{d_2,\{1,7\},3}\oplus f_{d_7,\{1,2\},1}$ to relay $2$ and user $1$ can use $f_{d_7,\{1,2\},1}$  and $f_{d_2,\{1,7\},3}$ to decode $f_{d_1,\{2,7\},4}$. On the other hand, our proposed scheme is based on the asymmetric coded placement in~\cite{Wan2018ITA} which is proved to be better than the scheme in~\cite{Zewail2017codedcaching}.
\end{example}
We now generalize the proposed scheme in Example~\ref{ex:cache-aided relays and users}.  Notice that in this example,  $f_{i,\{1,2\}}$ with $\Bsf/36$ bits is divided into $|\Rc_{\{1,2\}}|+1=3$ non-overlapping pieces where $|f_{i,\{1,2\},1}|=|f_{i,\{1,2\},2}|=\frac{\Msf^{\textrm{relay}}\Bsf}{\Nsf\binom{\Ksf_1}{g-1}}=\Bsf/72$ bits and $|f^{\prime}_{i,\{1,2\}}|=0$.
 It can be seen that if we increase $\Msf^{\textrm{relay}}$ by a small value and we still desire to achieve $g=3$, we have $|f_{i,\{1,2\},1}|+|f_{i,\{1,2\},2}|>f_{i,\{1,2\}}$ and thus these two pieces are overlapped which leads to redundancy. In other words, not all the bits of $F_{d_1}$ cached in relays $1$  and $2$ are useful to user $1$. 
So in this paper, we only consider the case 
\begin{align*}
&\Msf^{\textrm{relay}}\leq\frac{\Nsf \binom{\Ksf_1}{g-1}}{\max_{\Wc\in \Zc_{g}}|\Rc_{\Wc}| \Ssf_1(g)}\\
&=\frac{\Nsf \binom{\Ksf_1}{g-1}}{\max\{y\in [\rsf]:\Ksf_y\geq g-1\} \Ssf_1(g)},
\end{align*}
where $\Bsf/\Ssf_1(g)$ is the length of each MDS symbol generated by the scheme in~\cite{Wan2018ITA} (described in Section~\ref{sub:our previous scheme}).

The memories-loads tradeoff of the proposed scheme  is the lower convex envelope of the three groups of points.
\begin{enumerate}
\item $(\Msf^{\textrm{relay}},\Msf^{\textrm{user}})=\left(0,\frac
{\Nsf\Ssf_2(g)}
{\Ssf_1(g)}\right)$ for each $g\in [1:\Ksf_1]$. For each point in this group, we can see that relays do not have memory and the scheme is equivalent to the one for combination networks with end-user-caches. We use the scheme in Section~\ref{sub:our previous scheme} which leads $(\Rsf^{\ssf\to \rsf},\Rsf^{\rsf\to \usf})=\big(\Ksf(1-\Msf^{\textrm{user}}/\Nsf)/(\Hsf g),(1-\Msf^{\textrm{user}}/\Nsf)/\rsf\big).$
\item 
$(\Msf^{\textrm{relay}},\Msf^{\textrm{user}})=\Big(\frac{\Nsf \binom{\Ksf_1}{g-1}}{\max\{y\in [\rsf]:\Ksf_y\geq g-1\}\Ssf_1(g)},\Nsf\frac{\Ssf_2(g)}
{\Ssf_1(g)}$ $-\frac{\Nsf\binom{\Ksf_1-1}{g-2}\rsf}{\max\{y\in [\rsf]:\Ksf_y\geq g-1\} \Ssf_1(g)} \Big)$ where the coded caching gain $g\in [2:\Ksf_2+1]$.
In this case, for each user $k$, since each relay $h\in\Hc_k$ caches $f_{i,\Wc,h}$ where $k\in \Wc$, $|\Wc|=g-1$ and $\Wc\subseteq \Uc_h$, we have
\begin{align}
&\sum_{i\in [\Nsf]}\sum_{\Wc\in \Zc_{g}:k\in \Wc}\sum_{h\in \Hc_k:\Wc\subseteq \Uc_h}|f_{i,\Wc,h}|\nonumber\\
&=\sum_{i\in [\Nsf]}\sum_{h\in \Hc_k}\nonumber\sum_{\Wc\subseteq \Uc_h:k\in \Wc,|\Wc|=g-1}|f_{i,\Wc,h}|\nonumber\\
&=\frac{\Nsf\binom{\Ksf_1-1}{g-2}\rsf}{\max\{y\in [\rsf]:\Ksf_y\geq g-1\} \Ssf_1(g)}.\label{eq:relay caches}
\end{align}
In addition, since each user $k$ caches $f^{\prime}_{i,\Wc}$ where $k\in \Wc$, we have
\begin{align}
\Msf^{\textrm{user}}=\sum_{i\in [\Nsf]}\sum_{\Wc\in \Zc_{g}:k\in \Wc}|f^{\prime}_{i,\Wc}|\label{eq:user caches}.
\end{align}
Hence, from~\eqref{eq:relay caches} and~\eqref{eq:user caches} we have 
\begin{align*}
&\sum_{i\in [\Nsf]}\sum_{\Wc\in \Zc_{g}:k\in \Wc}|f^{\prime}_{i,\Wc}|+\sum_{i\in [\Nsf]}\sum_{\Wc\in \Zc_{g}:k\in \Wc}\sum_{h\in \Hc_k}|f_{i,\Wc,h}|\\
&=\Nsf\frac{\Ssf_2(g)}{\Ssf_1(g)}=\sum_{i\in [\Nsf]}\sum_{\Wc\in \Zc_{g}:k\in \Wc}|f_{i,\Wc}|.
\end{align*}
Hence, we can use  the proposed scheme in Example~\ref{ex:cache-aided relays and users}. 
\paragraph*{Placement phase}
We also divide each $F_{i}$ into $\Ssf_1(g)$ non-overlapping and equal-length pieces, which are then encoded by  $(|\Zc_{g}|,\Ssf_1(g))$ MDS code. The length of each MDS symbol is $\Bsf/\Ssf_1(g)$.
For each $\Wc\in \Zc_{g}$, there is one MDS symbol  denoted by $f_{i,\Wc}$ and we divide $f_{i,\Wc}$ into $|\Rc_{\Wc}|+1$ non-overlapping parts, $f_{i,\Wc}=\{f_{i,\Wc,h}:h\in \Rc_{\Wc}\}\cup\{f^{\prime}_{i,\Wc}\}$. For each $h\in \Rc_{\Wc}$, $f_{i,\Wc,h}$ is cached by relay $h$ where 
\begin{align*}
&|f_{i,\Wc,h}|\negmedspace=\negmedspace\frac{\Msf^{\textrm{relay}}\Bsf}{\Nsf\binom{\Ksf_1}{g-1}}\negmedspace=\negmedspace\frac{1}{\max\{y\in [\rsf]\negmedspace:\negmedspace\Ksf_y\geq g-1\}\Ssf_1(g)}.
\end{align*}
In addition, $f^{\prime}_{i,\Wc}$ is cached by each user in $\Wc$ where
\begin{align*}
&|f^{\prime}_{i,\Wc}|=\Bsf/\Ssf_1(g)-\frac{|\Rc_{\Wc}|\Msf^{\textrm{relay}}\Bsf}{\Nsf\binom{\Ksf_1}{g-1}}\\
&=\Bsf/\Ssf_1(g)-\frac{|\Rc_{\Wc}|\Bsf}{\Ssf_1(g)\max_{\Wc^{\prime}\in \Zc_{g}}|\Rc_{\Wc^{\prime}}|}.
\end{align*}
Hence, for each $\Wc\in \Zc_g$, if $\Wc\in \arg\max_{\Wc^{\prime}\in \Zc_{g}}|\Rc_{\Wc^{\prime}}|$, we have $|f^{\prime}_{i,\Wc}|=0$; otherwise, $|f^{\prime}_{i,\Wc}|>0$.

\paragraph*{Delivery phase}
We let each user $k$ recover $f_{d_k,\Wc}$ where $\Wc \in \Zc_{g}$ and $\Rc_{\{k\}\cup \Wc}\neq\emptyset$. There are two steps in delivery phase:
\begin{enumerate}
\item For each relay $h\in [\Hsf]$ and each user $k\in \Uc_h$, relay $h$ delivers all the cached bits of $F_{d_k}$ to user $k$.  More precisely, for each set $\Wc\subseteq \Uc_h$ where $|\Wc|=g-1$, relay $h$ delivers $f_{d_k,\Wc,h}$ to user $k$.

In addition, for each set $\Wc^{\prime}\subseteq \Uc_h$ where $|\Wc^{\prime}|=g-1$, $k\in \Wc^{\prime}$ and $|\Rc_{\Wc^{\prime}}|>1$, relay $h$ delivers $f_{d_{k^{\prime}},\Wc^{\prime},h}$ to user $k$ where $k^{\prime}\subseteq [\Ksf]\setminus \Uc_h$ and $\Rc_{\{k^{\prime}\}\cup \Wc^{\prime}}\neq \emptyset$. 
\item We let each user $k$ recover $f_{d_k,\Wc}\setminus\{f_{d_k,\Wc,h}:h\in (\Rc_{\Wc}\cap\Hc_k)\}$ where  $\Wc \in \Zc_{g}$, $k\notin \Wc$ and $\Rc_{\{k\}\cup\Wc}\neq \emptyset$. 
More precisely, we let $\Gc_{k,\Wc}=f_{d_k,\Wc}\setminus\{f_{d_k,\Wc,h}:h\in (\Rc_{\Wc}\cap\Hc_k)\}$ representing the unknown bits in $f_{d_k,\Wc}$ of user $k$.  We divide $\Gc_{k,\Wc}$ into $|\Rc_{\{k\}\cup\Wc}|$ non-overlapping and equal-length pieces, $\Gc_{k,\Wc}=\{\Gc_{k,\Wc,h}:h\in \Rc_{\{k\}\cup\Wc}\}$.
After considering all the MDS coded symbols demanded by all the  users, 
for each relay $h\in[\Hsf]$ and each set $\Jc\subseteq \Uc_h$ where $|\Jc|=g$, we create the multicast message in~\eqref{eq:proposed scheme multicast messages}, which is
to be sent to relay $h$ and then forwarded to the users in $\Jc$. It is also easily to check that each subfile in the multicast message in~\eqref{eq:proposed scheme multicast messages} has the same length.
\end{enumerate} 
We can compute that 
\begin{align*}
&\Rsf^{\ssf\to \rsf}=\frac{\Ksf\max \big\{1-(\rsf\Msf^{\textrm{relay}}+\Msf^{\textrm{user}})/\Nsf,0\big\}}{\Hsf g},\\
&\Rsf^{\rsf\to \usf}=\frac{(1-\Msf^{\textrm{user}}/\Nsf)}{\rsf}+\\
&\frac{\sum^{\rsf}_{b=2}\binom{\rsf-1}{b-1}\binom{\Ksf_b-1}{g-2}(\Ksf_1-\Ksf_2)(-1)^b}{\max\{y\in [\rsf]:\Ksf_y\geq g-1\}\Ssf_1(g)}.
\end{align*}

\item $(\Msf^{\textrm{relay}},\Msf^{\textrm{user}})=(\Nsf/\rsf,\Nsf t_1/\Ksf_1)$ where $t_1\in \{0,\Ksf_1\}$. In this case, each relay directly caches $s^h_{i}$ such that the server needs not to transmit any packets to relays. 

If $t_1=0$, each user does not cache any bits. In the delivery phase, each relay $h\in[\Hsf]$ transmits $s^h_{d_k}$ to each user $k\in\Uc_h$. So we have $(\Rsf^{\ssf\to \rsf},\Rsf^{\rsf\to \usf})=(0,1/\rsf)$.

If $t_1=\Ksf_1$, in the placement phase, each user $k$ caches $s^h_{i}$ for $h\in[\Hsf]$ and $i\in[\Nsf]$. So it caches all the $\Nsf$ files and $(\Rsf^{\ssf\to \rsf},\Rsf^{\rsf\to \usf})=(0,0)$. 
\end{enumerate}
\begin{rem}
\label{rem:improvement}
In~\cite{Wan2018ITA}, an improved scheme was proposed for combination network with end-user-caches when $g\in \big[\Ksf_3+2,\Ksf_2+\binom{\Hsf-3}{\rsf-2}\big]$. We can also use the above method to extend this improved scheme to combination networks with cache-aided relays and users. In this paper, for simplicity, we do not give the details.
\end{rem}

It is also straightforward to derive the following theorem.
\begin{thm}
\label{thm:comparison}
In a $(\Hsf,\rsf,\Msf^{\textrm{relay}},\Msf^{\textrm{user}},\Nsf)$ combination network, the achieved max link-load from the server to relays by the proposed scheme is not larger than the one achieved by the caching scheme in~\cite{Zewail2017codedcaching}.
\end{thm} 
\begin{figure}
\centerline{\includegraphics[scale=0.6]{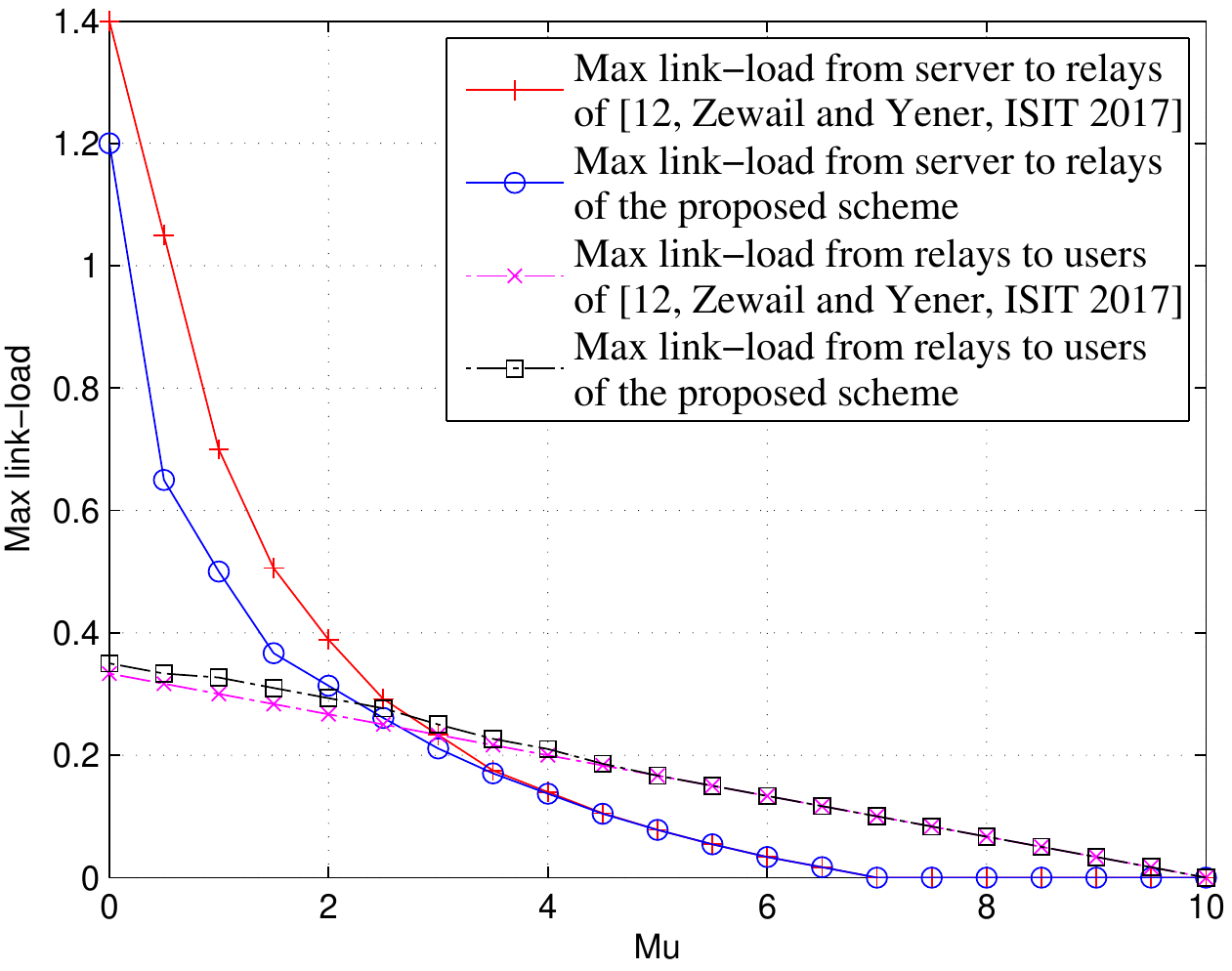}}
\caption{\small A combination network with cache-aided relays and users, with $\Hsf=5$ relays, $\rsf=2$,  $\Nsf=\Ksf=\binom{\Hsf}{\rsf}$ and $\Msf^{\textrm{relay}}=1$.}
\label{fig:rateh6r2}
\vspace{-5mm}
\end{figure}
\section{Numerical Comparisons and Conclusions}
\label{sec:numerical}
In Fig.~\ref{fig:rateh6r2} we compare the performance of the proposed scheme with the scheme in~\cite{Zewail2017codedcaching} for the centralized combination network with $\Hsf=5$, $\rsf=3$, and $\Nsf=\Ksf=10$ and $\Msf^{\textrm{relay}}=1$. 
We can see our proposed scheme provides a lower max link-load from the server to relays with the tradeoff of an increase of the max link-load of the max link-load from the relays to users. This is because that in the second step of delivery phase, each relay transmit some bits which are not from the demanded file of each of its connected users and these bits can lead an increased coded caching gain on the number of bits sent from the server.
 
To conclude the paper, we proposed a novel caching scheme for combination networks with cache-aided relays and users,  
 which aimed to create multicasting opportunities  from the caches of relays and users. 
The proposed scheme was shown to be achieve a max-link load from the server to relays not larger than the best scheme known in the litterature.

This work was supported in parts by NSF 1527059 and Labex DigiCosme.

\bibliographystyle{IEEEtran}
\bibliography{IEEEabrv,IEEEexample}
\end{document}